\documentclass[aps,prl,twocolumn,superscriptaddress]{revtex4}
\usepackage{graphicx}% Include figure files
\usepackage{dcolumn}% Align table columns on decimal point
\usepackage{bm}% bold math
\usepackage{verbatim}
\usepackage{color}
\usepackage[colorlinks]{hyperref}
\usepackage{amsmath}
\newcommand{\rttensor}[1]{\overline{\overline{#1}}}
\input{epsf}
\usepackage[sort&compress]{natbib}
\begin{document}

%\preprint{APS/123-QED}

\title{Magneto-transport characteristics of a 2D electron system driven to negative magneto-conductivity by microwave photoexcitation}

\author{R. G. Mani}
\affiliation{Department of Physics and Astronomy, Georgia State
University, Atlanta, GA 30303.}
\author{A. Kriisa}
\affiliation{Department of Physics, Emory University, 400 Dowman
Drive, Atlanta, GA 30322}
%\author{W. Wegscheider}
%\affiliation{Laboratorium f\"{u}r Festk\"{o}rperphysik, ETH
%Z\"{u}rich, 8093 Z\"{u}rich, Switzerland}

\date{\today}% It is always \today, today,

\begin{abstract}
Negative diagonal magneto-conductivity/resistivity is a
spectacular- and thought provoking- property of driven,
far-from-equilibrium, low dimensional electronic systems. The
physical response of this exotic electronic state is not yet fully
understood since it is rarely encountered in experiment. The
microwave-radiation-induced zero-resistance state in the high
mobility GaAs/AlGaAs 2D electron system is believed to be an
example where negative magneto-conductivity/resistivity is
responsible for the observed phenomena. Here, we examine the
magneto-transport characteristics of this negative
conductivity/resistivity state in the microwave photo-excited
two-dimensional electron system (2DES) through a numerical
solution of the associated boundary value problem. The results
suggest, surprisingly, that a bare negative diagonal
conductivity/resistivity state in the 2DES under photo-excitation
should yield a positive diagonal resistance with a concomitant
sign reversal in the Hall voltage.
\end{abstract}

%\pacs{72.20.My, 72.20.Fr, 72.80.Ey, 73.43.Fj }% PACS, the Physics and Astronomy
\maketitle

%\section{introduction}

Negative magneto-conductivity/resistivity is a spectacular- and
thought provoking- theoretical property of microwave photoexcited,
far-from-equilibrium, two-dimensional electronic systems. This
property has been utilized to understand the experimental
observation of the microwave-radiation-induced zero-resistance
states in the GaAs/AlGaAs system.\cite{grid1, grid2} Yet, the
negative conductivity/resistivity state remains an enigmatic and
open topic for investigation, although, over the past decade,
photo-excited transport has been the subject of a broad and
intense experimental \cite{ grid1, grid2, grid3, grid4, grid201,
grid202, grid5, grid6, grid7, grid8, grid9, grid11, grid209,
grid203, grid12, grid13, grid15, grid16, grid17, grid19, grid20,
grid16b, grid204,
 grid21z, grid21b, grid104, grid21, grid21y, grid22,
grid60, grid61, grid22z, grid22a, grid22c, grid108, grid109} and
theoretical\cite{grid23, grid24, grid25, grid101, grid27, grid28,
grid29, grid30, grid111, grid31, grid32, grid33, grid34, grid35,
grid37, grid206, grid39, grid40, grid42, grid43, grid44, grid45,
grid46, grid47, grid49, grid112, grid62, grid63, grid107, grid103}
study in the 2D electron system (2DES).

In experiment, the microwave-induced zero-resistance states arise
from "1/4-cycle-shifted" microwave radiation-induced
magnetoresistance oscillations in the high mobility GaAs/AlGaAs
system\cite{grid1, grid4, grid22a} as these oscillations become
larger in amplitude with the reduction of the temperature, $T$, at
a fixed microwave intensity. At sufficiently low $T$ under optimal
microwave intensity, the amplitude of the microwave-induced
magnetoresistance oscillations becomes large enough that the
deepest oscillatory minima approach zero-resistance. Further
reduction in $T$ then leads to the saturation of the resistance at
zero, leading to the zero-resistance states that empirically look
similar to the zero-resistance states observed under quantized
Hall effect conditions.\cite{grid1, grid2, grid5} Similar to the
situation in the quantized Hall effect, these radiation-induced
zero resistance states exhibit activated transport.\cite{grid1,
grid2, grid5, grid8} A difference with respect to the quantized
Hall situation, however, is that the Hall resistance, $R_{xy}$,
does not exhibit plateaus or quantization in this instance where
the zero-resistance state is obtained by
photo-excitation.\cite{grid1, grid2, grid5}

Some theories have utilized a two step approach to explain the
microwave-radiation-induced zero-resistance states. In the first
step, theory identifies a mechanism that helps to realize
oscillations in the diagonal
magneto-photo-conductivity/resistivity, and provides for the
possibility that the minima of the oscillatory diagonal
conductivity/resistivity can even take on negative
values.\cite{grid23, grid25, grid27, grid111, grid46, grid33}  The
next step in the two step approach invokes the theory of Andreev
et al.,\cite{grid24} who suggest that the zero-current-state at
negative resistivity (and conductivity) is unstable, and that this
favors the appearance of current domains with a  non-vanishing
current density,\cite{grid24, grid43} followed by the
experimentally observed zero-resistance states.

There exist alternate approaches which directly realize
zero-resistance states without a de-tour through negative
conductivity/resistivity states. Such theories include the
radiation-driven electron-orbit- model,\cite{grid34} the
radiation-induced-contact-carrier-accumulation/depletion
model,\cite{grid112} and the synchronization model.\cite{grid103}
Thus far, however, experiment has been unable to clarify the
underlying mechanism(s), so far as the zero-resistance states are
concerned.

The negative magneto-conductivity/resistivity state suggested
theoretically in this problem\cite{grid23, grid25, grid27,
grid111, grid46, grid33} has been a puzzle for experiment since it
had not been encountered before in magneto-transport. Naively, one
believes that negative magneto-resistivity/conductivity should
lead to observable negative magneto-resistance/conductance, based
on expectations for the zero-magnetic-field situation. At the same
time, one feels that the existence of the magnetic field is an
important additional feature, and this raises several questions:
Could the existence of the magnetic field be sufficiently
significant to overcome nominal expectations, based on the
zero-magnetic-field analogy, for an instability in a negative
magneto-conductivity/resistivity state? If an instability does
occur for the negative magneto-conductivity/resistivity state,
what is the reason for the instability? Could negative
conductivity/resistivity lead to observable negative
conductance/resistance at least in some short time-scale transient
situation where current domains have not yet formed? Indeed, one
might ask: what are the magneto-transport characteristics of a
bare negative conductivity/resistivity state? Remarkably, it turns
out that an answer has not yet been formulated for this last
question.

To address this last question, we examine here the transport
characteristics of the photo-excited 2DES at negative diagonal
conductivity/resistivity through a numerical solution of the
associated boundary value problem. The results suggest, rather
surprisingly, that negative conductivity/resistivity in the 2DES
under photo-excitation should generally yield a positive diagonal
resistance, i.e., $R_{xx} > 0$, except at singular points where
$R_{xx}=0$ when the diagonal conductivity $\sigma_{xx}=0$. The
simulations also identify an associated, unexpected sign reversal
in the Hall voltage under these conditions. These features suggest
that nominal expectations, based on the zero-magnetic-field
analogy, for a negative conductivity/resistivity state in a
non-zero magnetic field, need not necessarily follow, and that
experimental observations of zero-resistance and a linear Hall
effect in the photo-excited GaAs/AlGaAs system could be signatures
of vanishing conductivity/resistivity.

%\begin{comment}
\begin{figure}[t]
%h=here, t=top, b=bottom, p=separate figure page
\begin{center}
\leavevmode \epsfxsize=2.5 in \epsfbox {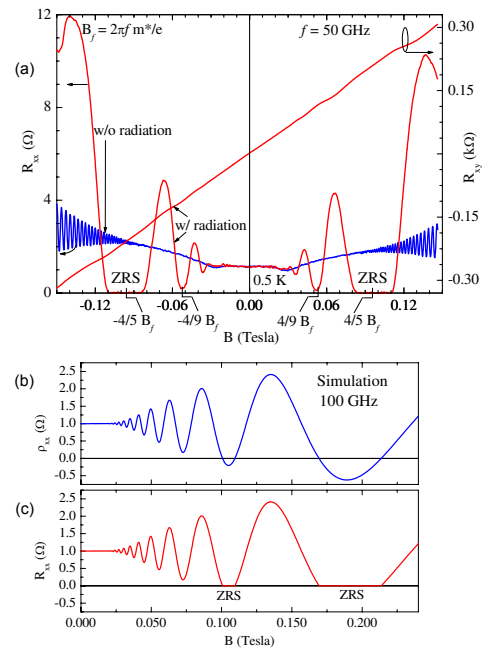}
\end{center}
%\begin{figure}
%\includegraphics{fig2}% Here is how to import EPS art
\caption{(Color online) (a) The dark- and photo-excited-diagonal-
($R_{xx}$) and the photo-excited-Hall- resistance ($R_{xy}$) are
exhibited vs. the magnetic field $B$ for a GaAs/AlGaAs
heterostructure at $T=0.5$K.  $R_{xx}$ exhibits a non-vanishing
resistance with Shubnikov-de Haas oscillations in the dark (blue
trace) at $|B| \ge 0.1$Tesla. Under photo-excitation at $f=50$GHz
(red traces), $R_{xx}$ exhibits large magnetoresistance
oscillations with vanishing resistance in the vicinity of $\pm
(4/5) B_{f}$, where $B_{f} = 2 \pi f m^{*}/e$. Note the absence of
a coincidental plateau in $R_{xy}$. (b) Theory predicts negative
diagonal resistivity, i.e., $\rho_{xx} < 0$, under intense
photoexcitation at the oscillatory minima, observable here in the
vicinity of $B \approx 0.19$ Tesla and $B \approx 0.105$ Tesla.
(c) Theory asserts that negative resistivity states are unstable
to current domain formation and zero-resistance. Consequently, the
$B$-span of negative resistivity in panel (b) corresponds to the
domain of zero-resistance states (ZRS), per theory. \label{fig:
epsart}}
\end{figure}

\begin{figure}[t]
%h=here, t=top, b=bottom, p=separate figure page
\begin{center}
\leavevmode \epsfxsize=3.25 in \epsfbox {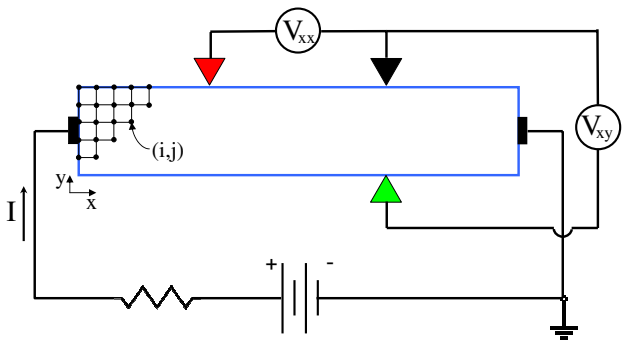}
\end{center}
%\begin{figure}
%\includegraphics{fig1}% Here is how to import EPS art
\caption{ (Color online) This dual purpose figure illustrates an
idealized measurement configuration and the simulation mesh. A
Hall bar (blue outline) is connected via its current contacts
(thick black rectangles at the ends) to a constant current source,
which may be modelled as a battery with a resistor in series. For
convenience, the negative pole of the battery has been grounded to
set the potential of this terminal to zero. A pair of "voltmeters"
are used to measure the diagonal $(V_{xx})$ and Hall $(V_{xy})$
voltages. For the numerical simulations reported in this work, the
Hall bar is represented by a mesh of points $(i,j)$, where the
potential is evaluated by a relaxation method. Here, $0 \le i \le
100$ and $0 \le j \le 20$. The long (short) axis of the Hall bar
corresponds the x (y)-direction. \label{afig: epsart}}
\end{figure}

\begin{figure}[t]
%h=here, t=top, b=bottom, p=separate figure page
\begin{center}
\leavevmode \epsfxsize=1.2 in \epsfbox {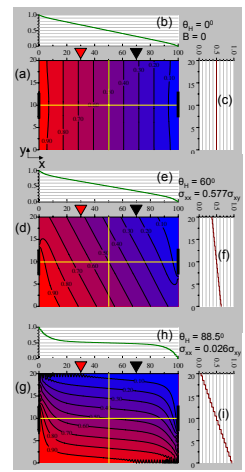}
\end{center}
%\begin{figure}
%\includegraphics{fig1}% Here is how to import EPS art
\caption{ (Color online) This figure summarizes the potential
profiles within a Hall bar device that is 100 units long and 20
units wide at three values of the Hall angle, $\theta_{H}$, where
$\theta_{H} = tan^{-1}(\sigma_{xy}/\sigma_{xx})$. (a) This panel
shows the potential profile at $\theta_{H} = 0^{0}$, which
corresponds to the $B=0$ situation. Current contacts are indicated
by black rectangles at the left- and right- ends, midway between
the top and the bottom edges. The left end of the Hall bar is at
$V = 1$ and the right end is at $V=0$. Potential (V) is indicated
in normalized arbitrary units. Panel (b) shows that the potential
decreases linearly along the indicated yellow line at $y=10$ from
the left end to the right end of the device. Panel (c) shows the
absence of a potential difference between the top- and bottom-
edges along the indicated yellow line at $x = 50$. That is, there
is no Hall effect at $B=0$. Panel (d) shows the potential profile
at $\theta_{H} = 60^{0}$, which corresponds to $\sigma_{xx} =
0.577 \sigma_{xy}$. Note that the equipotential contours develop a
tilt with respect to the same in panel (a). Panel (e) shows the
potential drop from the left to the right edge along the line at
$y=10$. Panel (f) shows a decrease in the potential from the
bottom to the top edge. This potential difference is the Hall
voltage at $\theta_{H} = 60^{0}$. Panel (g) shows the potential
profile at $\theta_{H} = 88.5^{0}$, which corresponds to
$\sigma_{xx} = 0.026 \sigma_{xy}$. Note that in the interior of
the device, the equipotential contours are nearly parallel to the
long axis of the Hall bar, in sharp contrast to (a). Panel(h)
shows the potential variation from the left to the right end of
the device along the line at $y=10$. The reduced potential
variation here between the $V_{xx}$ voltage probes (red and black
triangles) is indicative of a reduced diagonal resistance.
Panel(i) shows a large variation in the potential along the line
at $x=50$ between the bottom and top edges.  \label{afig2:
epsart}}
\end{figure}

%\section{Results}
\begin{figure}[t]
%h=here, t=top, b=bottom, p=separate figure page
\begin{center}
\leavevmode \epsfxsize=2 in \epsfbox {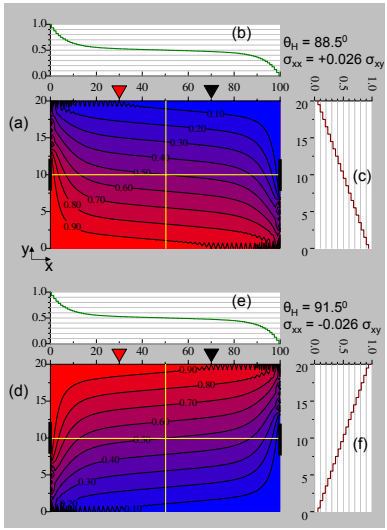}
\end{center}
%\begin{figure}
%\includegraphics{fig2}% Here is how to import EPS art
\caption{(Color online) This figure compares the potential profile
within the Hall bar device for positive ($\sigma_{xx}= +0.026
\sigma_{xy}$) and negative ($\sigma_{xx}= -0.026 \sigma_{xy}$)
conductivity. Panel (a) shows the potential profile at $\theta_{H}
= 88.5^{0}$ which corresponds to $\sigma_{xx}= +0.026
\sigma_{xy}$. Note that in the interior of the device, the
equipotential contours are nearly parallel to the long axis of the
Hall bar. Panel(b) shows the potential variation from the left to
the right end of the device along the line at $y=10$. The small
potential drop here between the $V_{xx}$ voltage probes (red and
black triangles) is indicative of a reduced diagonal resistance in
this low $\sigma_{xx}$ condition. Panel(c) suggests the
development of a large Hall voltage between the bottom and top
edges. Here the voltage decreases towards the top edge. Panel (d)
shows the potential profile at $\theta_{H} = 91.5^{0}$ which
corresponds to $\sigma_{xx}= -0.026 \sigma_{xy}$. The key feature
here is the reflection of the potential profile with respect panel
(a) about the line at $y=10$ when the $\sigma_{xx}$ shifts from a
positive ($\sigma_{xx}= +0.026 \sigma_{xy}$) to a negative
($\sigma_{xx}= -0.026 \sigma_{xy}$) value. Panel (e) shows that in
the negative $\sigma_{xx}$ condition, the potential still
decreases from left to right, implying a positive diagonal voltage
$V_{xx}$ and diagonal resistance $R_{xx}$. Panel (f) shows that
for $\sigma_{xx}= -0.026 \sigma_{xy}$, the potential
\textit{increases} from the bottom edge to the top edge, unlike in
panel (c). Thus, the Hall voltage undergoes sign reversal in going
from the $\sigma_{xx}= +0.026 \sigma_{xy}$ situation to the
$\sigma_{xx}= -0.026 \sigma_{xy}$ condition, compare panels (c)
and (f).\label{fig3:epsart}}
\end{figure}

\begin{figure}[t]
%h=here, t=top, b=bottom, p=separate figure page
\begin{center}
\leavevmode \epsfxsize=2.25 in \epsfbox {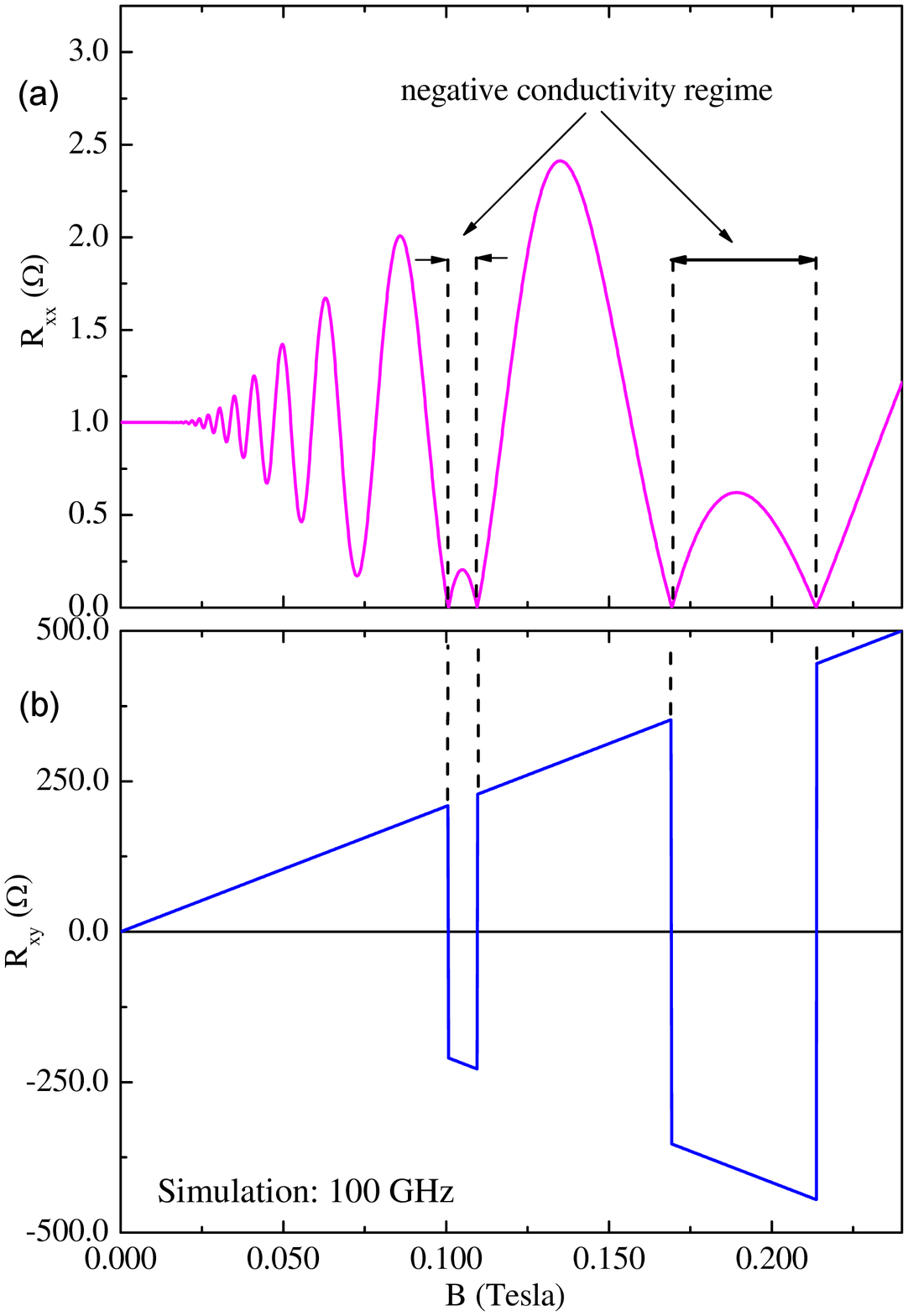}
\end{center}
%\begin{figure}
%\includegraphics{fig2}% Here is how to import EPS art
\caption{ (Color online) This figure illustrates expectations,
based on the results illustrated in Fig. 3 and 4, for the behavior
of the diagonal ($R_{xx}$) and Hall ($R_{xy}$) resistances in a 2D
system driven periodically to negative conductivity/resistivity by
photo-excitation. (a) The diagonal resistance $R_{xx}$ exhibits
microwave-induced magnetoresistance oscillations that grow in
amplitude with increasing $B$. In the regime of negative
conductivity at the oscillatory minima, the $R_{xx}$ exhibits
positive values. (b) Over the same span of $B$, the Hall
resistance $R_{xy}$ shows sign reversal.  \label{fig:epsart}}
\end{figure}
%\end{comment}
\section{Results}
\subsection{Experiment}
Figure 1(a) exhibits measurements of $R_{xx}$ and $R_{xy}$ over
the magnetic field span $-0.15 \le B \le 0.15$ Tesla at $T=0.5K$.
The blue curve, which exhibits Shubnikov-de Haas oscillations at
$|B| \ge 0.1$ Tesla, represents the $R_{xx}$ in the absence of
photo-excitation (w/o radiation). Microwave photo-excitation of
this GaAs/AlGaAs specimen at $50$GHz, see red traces in Fig. 1,
produces radiation-induced magnetoresistance oscillations in
$R_{xx}$, and these oscillations grow in amplitude with increasing
$|B|$. At the deepest minimum, near $|B| = (4/5)B_{f}$, where
$B_{f} = 2 \pi f m^{*}/e$,\cite{grid1} the $R_{xx}$ saturates at
zero-resistance. Note also the close approach to zero-resistance
near $|B| = (4/9) B_{f}$. Although $R_{xx}$ exhibits
zero-resistance, the Hall resistance $R_{xy}$ exhibits a linear
variation over the $B$-span of the zero-resistance states, see
Fig. 1(a).\cite{grid1,grid2}
\subsection{Negative magneto-resistivity and zero-resistance}
Both the displacement theory for the radiation-induced
magnetoresistivity oscillations,\cite{grid23} and the inelastic
model,\cite{grid33} suggest that the magnetoresistivity can take
on negative values over the $B$-spans where experiment indicates
zero-resistance states. For illustrative purposes, such
theoretical expectations for negative resistivity are sketched in
Fig. 1(b), which presents the simulated $\rho_{xx}$ at $f=100$
GHz. This curve was obtained on the basis of extrapolating,
without placing a lower bound, the results of fits,\cite{grid202}
which have suggested that the radiation-induced oscillatory
magnetoresistivity, $\rho_{xx}^{osc}$, where $\rho_{xx}^{osc}=
R_{xx}^{osc}(W/L)$, with $W/L$ the device width-to-length ratio,
follows $\rho_{xx}^{osc} = A'' exp(-\lambda/B) sin(2 \pi F/B -
\pi)$. Here, $F = 2 \pi f m^{*}/e$, with $f=100$GHz, the microwave
frequency, $m^{*} = 0.065m_{e}$, the effective mass, $e$, the
electron charge, and $\rho_{xx} = \rho_{xx}^{dark} +
\rho_{xx}^{osc}$, with $\rho_{xx}^{dark}$ the dark resistivity
which reflects typical material characteristics for the high
mobility GaAs/AlGaAs 2DES. This figure shows that the deepest
$\rho_{xx}$ minima at $B \approx 0.19$ Tesla and $B \approx 0.105$
Tesla exhibit negative resistivity, similar to theoretical
predictions.\cite{grid23, grid25, grid27, grid111, grid46, grid33}

Andreev et al.\cite{grid24} have reasoned that the only
time-independent state of a system with negative
resistivity/conductivity is characterized by a current which
almost everywhere has a magnitude $j_{0}$ fixed by the condition
that nonlinear dissipative resistivity equals zero. This
prediction implies that the $\rho_{xx}$ curve of Fig. 1(b) is
transformed into the magnetoresistance, $R_{xx}$, trace shown in
Fig. 1(c), where the striking feature is the zero-resistance over
the $B$-domains that exhibited negative resistivity in Fig. 1(b).
The curve of Fig. 1(c) follows from Fig. 1(b) upon multiplying the
ordinate by the $L/W$ ratio, i.e., $R_{xx} = \rho_{xx} (L/W)$, and
placing a lower bound of zero on the resulting $R_{xx}$.

\subsection{Device configuration}
As mentioned, a question of interest is: what are the transport
characteristics of a bare negative
magneto-conductivity/resistivity state? To address this issue, we
reexamine the experimental measurement configuration in Fig. 2.
Transport measurements are often carried out in the Hall bar
geometry which includes finite size current contacts at the ends
of the device. Here, a constant current is injected via the ends
of the device, and "voltmeters" measure the diagonal ($V_{xx}$)
and Hall ($V_{xy}$) voltages between probe points as a function of
a transverse magnetic field, as indicated in Fig. 2.
Operationally, the resistances relate to the measured voltages by
$R_{xx} = V_{xx}/I$ and $R_{xy}=V_{xy}/I$.
\subsection{Simulations}
Hall effect devices can be numerically simulated on a
grid/mesh,\cite{grid99,grid110,grid102} see Fig. 2, by solving the
boundary value problem corresponding to enforcing the local
requirement $\nabla . \overrightarrow{j} = 0$, where
$\overrightarrow{j}$ is the 2D current density with components
$j_{x}$ and $j_{y}$, $\overrightarrow{j} = \rttensor{\sigma}
\overrightarrow{E}$, and $\rttensor{\sigma}$ is the conductivity
tensor.\cite{grid99,grid110} Enforcing $\nabla .
\overrightarrow{j} = 0$ within the homogeneous device is
equivalent to solving the Laplace equation $\nabla^{2} V = 0$,
which may be carried out in finite difference form using a
relaxation method, subject to the boundary conditions that current
injected via current contacts is confined to flow within the
conductor. That is, current perpendicular to edges must vanish
everywhere along the boundary except at the current contacts. We
have carried out simulations using a $101 \times 21$ point grid
with current contacts at the ends that were 6 points wide. For the
sake of simplicity, the negative current contact is set to ground
potential, i.e., $V=0$, while the positive current contact is set
to $V=1$. In the actual Hall bar device used in experiment, the
potential at the positive current contact will vary with the
magnetic field but one can always normalize this value to 1 to
compare with these simulations.

Figure 3 summarizes the potential profile within the Hall device
at three values of the Hall angle, $\theta_{H} = tan^{-1}
(\sigma_{xy}/\sigma_{xx})$. Fig. 3(a) shows a color plot of the
potential profile with equipotential contours within the device at
$\theta_{H} = 0^{0}$, which corresponds to the $B=0$ situation.
This panel, in conjunction with Fig. 3(b), shows that the
potential drops uniformly within the device from the left- to the
right- ends of the Hall bar. Fig. 3(c) shows the absence of a
potential difference between the top- and bottom- edges along the
indicated yellow line at $x = 50$. This feature indicates that
there is no Hall effect in this device at $B=0$, as expected.

Figure 3(d) shows the potential profile at $\theta_{H} = 60^{0}$,
which corresponds to the situation where $\sigma_{xx} = 0.577
\sigma_{xy}$. Note that, here, the equipotential contours develop
a tilt with respect to the same in Fig. 3(a).
Fig. 3(e) shows a mostly uniform potential drop from the left to
the right edge along the line at $y=10$, as Fig. 3(f) shows a
decrease in the potential from the bottom to the top edge. This
potential difference represents the Hall voltage under these
conditions.

Figure 3(g) shows the potential profile at $\theta_{H} =
88.5^{0}$, which corresponds to the situation where $\sigma_{xx} =
0.026 \sigma_{xy}$. Note that in the interior of the device, the
equipotential contours are nearly parallel to the long axis of the
Hall bar, in sharp contrast to Fig. 3(a). Fig. 3(h) shows the
potential variation from the left to the right end of the device. The reduced change in potential
between the $V_{xx}$ voltage probes (red and black inverted triangles), in
comparison to Fig. 3(b) and Fig. 3(e) is indicative of a reduced
diagonal voltage and resistance. Fig. 3(i) shows a large potential difference  between the bottom and
top edges, indicative of a large Hall voltage.

The results presented in Fig. 3  display the normal expected
behavior for a 2D Hall effect device with increasing Hall angle.
Such simulations can also be utilized to examine the influence of
microwave excitation since microwaves modify the diagonal
conductivity, $\sigma_{xx}$, or resistivity,
$\rho_{xx}$,\cite{grid23, grid33} and this sets $\theta_{H}$ via
$\theta_{H} = tan^{-1} (\sigma_{xy}/\sigma_{xx})$. In the next
figure, we examine the results of such simulations when the
diagonal conductivity, $\sigma_{xx}$, reverses signs and takes on
negative values, as per theory, under microwave excitation. Thus,
figure 4 compares the potential profile within the Hall bar device
for positive ($\sigma_{xx}= +0.026 \sigma_{xy}$) and negative
($\sigma_{xx}= -0.026 \sigma_{xy}$) values of the conductivity.

Fig. 4(a) shows the potential profile at  $\sigma_{xx}= +0.026
\sigma_{xy}$. This figure is identical to the figure exhibited in
Fig. 3(g). The essential features are that the equipotential
contours are nearly parallel to the long axis of the Hall bar, see
Fig. 4(b), signifying a reduced diagonal resistance. Concurrently,
Fig. 4(c) suggests the development of a large Hall voltage between
the bottom and top edges. Here the Hall voltage decreases from the
bottom- to the top- edge.

Fig. 4(d) shows the potential profile at $\sigma_{xx}= -0.026
\sigma_{xy}$, i.e., the negative conductivity case. The important
feature here is the reflection of the potential profile with
respect Fig. 4(a) about the line at $y=10$ when the $\sigma_{xx}$
shifts from a positive ($\sigma_{xx}= +0.026 \sigma_{xy}$) to a
negative ($\sigma_{xx}= -0.026 \sigma_{xy}$) value. Fig. 4(e)
shows, remarkably, that in the negative $\sigma_{xx}$ condition,
the potential still decreases from left to right, implying
$V_{xx}>0$ and $R_{xx}>0$ even in this $\sigma_{xx} \le 0$
condition. Fig. 4(f) shows that for $\sigma_{xx}= -0.026
\sigma_{xy}$, the potential \textit{increases} from the bottom
edge to the top edge, in sharp contrast to Fig. 4(c). Thus, these
simulations show clearly that the Hall voltage undergoes sign
reversal when $\sigma_{xx} \le 0$, although the diagonal voltage
(and resistance) exhibits positive values.

\section{Discussion}
Existing theory indicates that photo-excitation of the high
mobility 2D electron system can drive the $\rho_{xx}$ and
$\sigma_{xx}$ to negative values at the minima of the
radiation-induced oscillatory magneto-resistivity.\cite{grid23,
grid25, grid27, grid111, grid46, grid33} Andreev et
al.,\cite{grid24} have argued that "$\sigma_{xx} < 0$ by itself
suffices to explain the zero-dc-resistance state" because
"negative linear response conductance implies that the
zero-current state is intrinsically unstable." Since our
simulations (Fig. 4) show clearly that negative magneto
conductivity/resistivity leads to positive, not negative,
conductance/resistance, it looks like one cannot argue for an
instability in the zero-current state based on presumed "negative
linear response conductance."

For illustrative purposes, using the understanding obtained from
the simulation results shown in Fig. 4, we sketch in Fig. 5 the
straightforward expectations,  for the behavior of the diagonal
($R_{xx}$) and Hall ($R_{xy}$) resistances in a 2D system driven
periodically to negative diagonal conductivity by
photo-excitation. Fig. 5(a) shows that the microwave-induced
magnetoresistance oscillations in $R_{xx}$ grow in amplitude with
increasing $B$. When the oscillations in the
magneto-resistivity/conductivity are so large that the oscillatory
minima would be expected to cross into the regime of $\sigma_{xx}
<0$ at the oscillatory minima, the $R_{xx}$ exhibits positive
values. Here, vanishing $R_{xx}$ occurs only at singular values of
the magnetic field where $\sigma_{xx} = 0$. Fig. 5(b) shows that
the Hall resistance $R_{xy}$ shows sign reversal over the same
span of $B$ where $\sigma_{xx} <0$.

It appears that if there were an instability, it should be related
to the sign-reversal in the Hall effect. Yet, note that sign
reversal in the Hall effect is not a manifestly un-physical effect
since it is possible to realize Hall effect sign reversal in
experiment even with a fixed external bias on the sample, as in
the simulations, simply by reversing the direction of the magnetic
field or by changing the sign of the charge carriers. The unusual
characteristic indicated by these simulations is Hall effect sign
reversal even without changing the direction of the magnetic field
or changing the sign of the charge carriers. This feature can be
explained, however, by noting that the numerical solution of the
boundary value problem depends on a single parameter, the Hall
angle, $\theta_{H}$, where $tan (\theta_{H}) =
\sigma_{xy}/\sigma_{xx}$. Since this single parameter depends on
the ratio of the off-diagonal and diagonal conductivities, sign
change in $\sigma_{xx}$ produces the same physical effect as sign
reversal in $\sigma_{xy}$ so far as the solution to the boundary
value problem is concerned. That is, one might change the sign of
$\sigma_{xy}$ or one might change the sign of $\sigma_{xx}$, the
end physical result is the same: a sign reversal in the Hall
effect.

One might also ask: why do the simulations indicate a positive
diagonal resistances for the negative diagonal
conductivity/resistivity scenario? The experimental setup shown in
Fig. 2 offers an answer to this question: In the experimental
setup, the Hall bar is connected to an external battery which
enforces the direction of the potential drop from one end of the
specimen to the other. This directionality in potential drop is
also reflected in the boundary value problem. As a consequence,
the red potential probe in Fig. 2, 3 or 4 would prefer to show a
higher potential than the black potential probe so long as the
$\sigma_{xx}$ is not identically zero, and this leads to the
positive resistance even for negative diagonal
conductivity/resistivity in the numerical simulations.

We remark that the experimental results of Fig. 1(a) are quite
unlike the expectations exhibited in Fig. 5. Experiment shows an
ordinary Hall effect without anomalies over the zero-resistance
region about $(4/5)B_{f}$, not a sign reversal in the Hall effect,
and experiment shows zero-resistance, not the positive resistance
expected for a system driven to negative conductivity.

In conclusion, the results presented here suggest that a bare
negative magneto- conductivity/resistivity state in the 2DES under
photo-excitation should yield a positive diagonal resistance with
a concomitant sign reversal in the Hall effect.

We have also understood that these results could be potentially
useful for understanding plateau formation in the Hall effect as,
for example, in the quantum Hall situation, if new physics comes
into play in precluding sign reversal in the Hall effect, when the
diagonal magneto-conductivity/resistivity is forced into the
regime of negative values.

\section{Methods}
\subsection{Samples}
The GaAs/AlGaAs material utilized in our experiments exhibit
electron mobility $\mu \approx 10^{7} cm^{2}/Vs$ and electron
density in the range $2.4 \times 10^{11} \leq n \leq 3 \times
10^{11} cm^{-2}$. Utilized devices include cleaved specimens with
alloyed indium contacts and Hall bars fabricated by optical
lithography with alloyed Au-Ge/Ni contacts. Standard low frequency
lock-in techniques yield the electrical measurements of $R_{xx}$
and $R_{xy}$.\cite{grid1,grid3,grid4,grid201, grid202,
grid5,grid11, grid209, grid203, grid16, grid16b, grid21, grid21y,
grid60, grid61, grid108, grid109}

\subsection{Microwave transport measurements}
Typically, a Hall bar specimen was mounted at the end of a long
straight section of a rectangular microwave waveguide. The
waveguide with sample was inserted into the bore of a
superconducting solenoid, immersed in pumped liquid Helium, and
irradiated with microwaves, at a source-power $0.1 \le P \le 10$
mW, as in the usual microwave-irradiated transport
experiment.\cite{grid1}  The applied external magnetic field was
oriented along the solenoid and waveguide axis.

%\section{Discussion}

%\section{Summary/Conclusion}

%\section{Acknowledgements}

%\pagebreak

\section{Acknowledgements}

The basic research at Georgia State University is primarily
supported by the U.S. Department of Energy, Office of Basic Energy
Sciences, Material Sciences and Engineering Division under
DE-SC0001762. Additional support is provided by the ARO under
W911NF-07-01-015.
\pagebreak
%\begin{comment}
%\begin{comment}
%%%%%%%%%%%%%%%%%%%%%%%%%%%%%%%%%%%%%%%%

%\pagebreak

%\pagebreak

%\pagebreak

%\end{comment}
%%%%%%%%%%%%%%%%%%%%%%%%%%%%%%%%%%%%%%%%

\end{document}